\documentclass[12pt]{elsarticle}

\usepackage{amsthm}
\usepackage{amssymb}
\usepackage{amsmath}
\usepackage{pdfpages}
\usepackage{wasysym}
\usepackage{txfonts}
\usepackage{longtable}
\usepackage{multirow}
\usepackage{float}
\usepackage{graphicx}
\usepackage{rotating}
\usepackage{pdflscape}
\usepackage{afterpage}
\usepackage{bm}
\usepackage{xcolor}
\usepackage{subcaption}

\begin{document}

\begin{frontmatter}



\title{Influence of Octahedral Site Chemistry on the Elastic Properties of Biotite}

\author[label1]{Dillon F. Hanlon}
\author[label1]{G. Todd Andrews}
\author[label2]{Roger A. Mason}


\address[label1]{Department of Physics and Physical Oceanography, Memorial University, St. John's, NL, Canada, A1B 3X7}
\address[label2]{Department of Earth Sciences, Memorial University, St. John's, NL, Canada, A1B 3X5}

\begin{abstract}
 Brillouin light scattering spectroscopy was used along with detailed composition information obtained from electron probe microanalysis to study the influence of octahedral site chemistry on the elastic properties of natural biotite crystals.  Elastic wave velocities for a range of directions in the $ac$ and $bc$ crystallographic planes were obtained for each crystal by application of the well-known Brillouin equation with refractive indices and phonon frequencies obtained from the Becke line test and spectral peak positions, respectively.  In general, these velocities increase with decreasing iron content, approaching those of muscovite at low iron concentrations. Twelve of thirteen elastic constants for the full monoclinic symmetry were obtained for each crystal by fitting analytic expressions for the velocities as functions of propagation direction and elastic constants to corresponding experimental data, while the remaining constant was estimated under the approximation of hexagonal symmetry.  Elastic constants $C_{11}$, $C_{22}$, and $C_{66}$ are comparable to those of muscovite and show little change with iron concentration due to the strong bonding within layers.  In contrast, nearly all of the remaining constants show a pronounced dependence on iron content, a probable consequence of the weak interlayer bonding.  Similar behaviour is displayed by the elastic stability, which exhibits a dramatic decrease with increasing iron content, and by the elastic anisotropy within the basal cleavage plane, which decreases as the amount of iron in the crystal is reduced.  This systematic dependence on iron content of all measured elastic properties indicates that the elasticity of biotite is a function of octahedral site chemistry and provides a means to estimate the elastic constants and relative elastic stability of most natural biotite compositions if the iron or, equivalently, magnesium, concentration is known.  Moreover, the good agreement between the elastic constants of Fe-poor (Mg-rich) biotite and those of phlogopite obtained from \textit{first-principles} calculation based on density functional theory indicates that the latter approach may be of use in predicting the elastic properties of biotites.
\end{abstract}


\begin{highlights}
\item Elastic properties of natural biotite crystals are functions of octahedral site chemistry.
\item Elastic stability of biotite shows dramatic decrease with increasing iron content.
\item Elastic properties and elastic stability of a biotite can be estimated if iron content is known.
\end{highlights}

\begin{keyword}
Biotite \sep Elastic Properties \sep Octahedral Site Chemistry \sep Brillouin light scattering \sep Electron Probe Microanalysis


\end{keyword}

\end{frontmatter}


\section{\label{sec:Intro}Introduction}
Micas are common rock-forming phyllosilicate minerals displaying monoclinic symmetry and comprising approximately 5-12\% of the continental crust \cite{almq2017}.  The perfect \{001\}
cleavage and platy habit of the mica group lead to preferred orientation of their grains.  In sedimentary rocks preferred
orientation arises as a depositional or diagenetic feature.  In
shales, composed predominantly of clay minerals, which are closely
related in structure to micas, such preferred orientation leads to
their fissility.  In metamorphic rocks preferred orientation of micas
(and other minerals) is a response to crystallisation (or
recrystallisation) in a stress field, causing the development of a
foliation (or schistosity) perpendicular to the direction of the
primary stress.  At the highest temperatures of metamorphism, and in
the presence of an anisotropic stress field, compositional
segregation, combined with foliation, may lead to the development of
very strongly layered gneisses (gneissosity).
 
The presence of fissility, foliation, or gneissosity causes anisotropy
in the velocity of seismic waves through the Earth.  This was
recognised in the nineteenth century and with the development of
improved instrumentation and processing techniques has become a tool
that is used both in exploration geophysics and deep Earth studies \cite{helb2005, roma2017, chan2021}.

The anisotropy of rocks can provide information on rock microstructure and crystal anisotropy bears on efforts to use inclusions of one mineral in another to obtain measures of pressure at the time of inclusion \cite{hea20}. Complete and accurate characterization of the elastic properties of rocks and the minerals
that compose them is therefore important to understanding the
composition and structure of the Earth.  On a microscopic scale,
knowledge of the elasticity of sheet silicates like micas provides
insight into the nature of bonding and acoustic wave behaviour in
layered materials.  Moreover, the use of micas in applications such as
flexible electronic devices also requires that the elastic properties
be known \cite{ma2016van,xu2018flexible}.

\subsection{The Micas}
\label{sec:Micas}

The silicates are, volumetrically, the most important mineral group in
the Earth's crust.  Their structures are based on tetrahedra,
{\it i.e.} sites in which each central cation (commonly Si or Al) is
coordinated by four oxygen atoms arrayed at the apices of a tetrahedron.
Sharing of the apical oxygens between adjacent tetrahedra, together
with substitution or larger cations on sites that arise between
tetrahedra give rise to a wide variety of silicate structures and
symmetries.

\begin{figure}
\centering
\includegraphics[scale = 0.4]{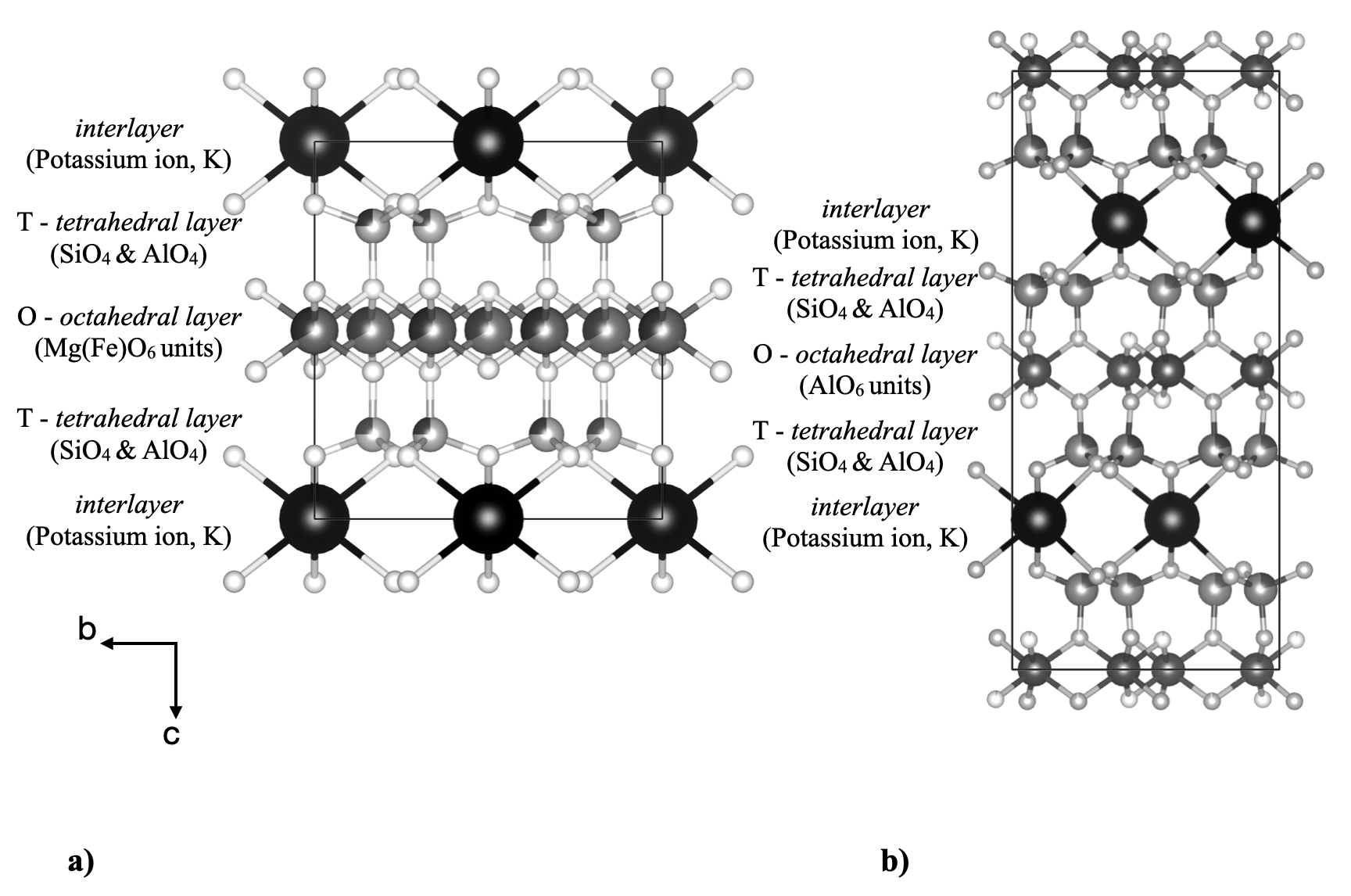}
\caption{Crystal structures of biotite and muscovite. \textit{a}) Biotite structure \cite{takeda1975mica} with two $T$ = Si, Al tetrahedral layer shown with light grey spheres which sandwich an octahedral layer. Mg (or Fe) octahedral layer shown with medium grey spheres. Interlayer potassium cations shown with black spheres.  b) Muscovite structure \cite{guggenheim1987muscovite} with interlayer cations shown with black spheres, tetrahedral cations shown with light grey spheres, and octahedral cations shown with medium grey spheres. }
\label{fig:micastruct}
\end{figure}

A key determinant of structure type is the number of oxygens per
tetrahedron that are shared with adjacent tetrahedra.  In the sheet
silicates three of the four oxygen atoms of each tetrahedral site are
shared, giving rise to the (Si, Al)-O tetrahedral sheet as the common
structural characteristic of these minerals.  Within the sheet
silicates the micas form a large and important group.  Their
distinguishing feature is a structure based on a double sheet of
tetrahedra between which is sandwiched a layer of octahedral sites
(coordination number six, see Fig. \ref{fig:micastruct}) in which four of
the six apices of each octahedron are coordinated to the oxygens of
the tetrahedra and two to sites that can be occupied by hydroxyl,
fluorine or chlorine.  There are two tetrahedral sheets per octahedral
sheet and the combined tetrahedral-octahedral-tetrahedral layer is
often called a TOT sheet.  The TOT sheets carry a net negative charge
that is compensated by large cations (K, Na, {\it etc}.) lying between them.
Various stacking sequences of the TOT sheets are possible but the most
common gives rise to monoclinic symmetry. Figure 1 was created using VESTA \cite{momma2011vesta} software along with data publicly available through the American Mineralogist database \cite{downs2003american} for biotite \cite{takeda1975mica} and muscovite \cite{guggenheim1987muscovite}.

\subsection{Biotite and Muscovite}
\label{sec:biomusc}

Biotite (K$_2$(Mg,Fe)$_6$[Si$_6$Al$_2$O$_{20}$](OH,F,Cl)$_4$) is trioctahedral
with three octahedral sites per formula unit (six in the conventional
unit cell, which we use here) completely or almost completely filled
(see left panel of Fig. \ref{fig:micastruct}).  Muscovite (K$_2$Al$_4$[Si$_6$Al$_2$O$_{20}$](OH,F,Cl)$_4$) is
dioctahedral, meaning that one of every three octahedra is vacant
(see right panel of Fig. \ref{fig:micastruct}).  In biotite, substitution of trivalent
(Fe\textsuperscript{3+}, Al\textsuperscript{3+}) or quadrivalent
(Ti\textsuperscript{4+}) cations on octahedral sites can be
compensated by either octahedral vacancies or by replacement of Si by
Al on tetrahedral sites beyond the ideal one out of four and there can
be extensive solid solution towards siderophyllite (an end member with
Fe and Al mixing on octahedral sites).  Similarly, in muscovite
altervalent substitution on the octahedral sites can be compensated by
changing the composition of the tetrahedra, although such
substitutions are less common than in biotite.  The chemical diversity
of the micas accounts for their widespread occurrence in igneous,
metamorphic and sedimentary rocks of widely differing chemistry and
paragenesis.

\subsection{Elasticity of Micas}
\label{sec:elasmicas}

Despite its importance and utility, the elasticity of many micas remains merely estimated or unexplored. This is largely due to chemical variability and the low symmetry of these
crystals. While complete sets of elastic constants have been measured
for muscovite \cite{vaug1986,mcne1993}, analogous data for the biotite
group have not been reported. Experimentally-determined elastic
constants for biotite are limited to estimates obtained via ultrasonic
techniques for two phlogopite samples and a biotite sample of
unspecified composition with the symmetry approximated as hexagonal
rather than the true monoclinic \cite{alek1961}. This precludes direct
comparison with theoretical results for the full monoclinic elastic
constants tensor for phlogopite \cite{chhe2014} and muscovite
\cite{militzer2011first}, but differences in some constants among
these samples hint at the dependence of biotite elasticity on
composition.  This effect has not, however, yet been quantified.

In this paper, the results of Brillouin light scattering experiments
on natural crystals of biotite are reported.  The crystals were
selected {\it a priori} to be compositionally different, with the
composition later quantified by electron probe microanalysis.
Directional dependences of elastic wave velocities in the $ac$ and
$bc$ crystal planes were measured from Brillouin peak frequency shifts
and refractive indices determined from the Becke line test. These
velocities, together with sample densities, permitted twelve of
thirteen elastic constants and related mechanical properties to be
determined for each of the samples. The results show that the elastic
properties of biotite depend on octahedral site chemistry.  Comparison
is also made to new and published results for muscovite and to elastic
constants of phlogopite determined from \textit{first-principles}
calculations.

\section{\label{sec:Theory}Theory}
\subsection{Brillouin Light Scattering}
Brillouin spectroscopy is a technique used to probe thermally-excited acoustic phonons (elastic waves) in a medium via the inelastic scattering of light. For a $180^\circ$ backscattering geometry such as that used here, conservation of energy and momentum applied to the scattering process  \cite{dil1982,spez2014} yield the phonon velocity
\begin{equation}
v = \frac{\Omega}{q}= \frac{f\lambda_i}{2n}.
\label{brillouineqn}
\end{equation}
Here, $f= \Omega/2\pi = |f_s - f_i|$ is the Brillouin peak frequency shift and $q = 4\pi n/\lambda_i$ is the magnitude of the wavevector of the probed phonon, where $n$ is the refractive index of the target material and $\lambda_{i}$ the wavelength of the incident light.

\subsection{Elastic Waves} \label{elasticwaves}
Acoustic modes may be considered as sound waves in a crystal since their wavelengths are much greater than the primitive unit cell dimensions. The equation that describes the motion of these waves is

\begin{equation}
\rho \frac{\partial ^2 u_i}{\partial t^2} = C_{ijkl} \frac{\partial ^2 u_k}{\partial x_j \partial x_l},
\label{eq:eom}
\end{equation}
where $\rho$ is the density of the medium, $C_{ijkl}$ is the elastic stiffness tensor, $u_i(x_i,t)$ is the particle displacement as a function of position, $x_i$, and time, $t$, with $i,j,k,l = 1, 2, 3$ \cite{haye2012,every1980general}.

Assuming plane wave solutions to Eq. \ref{eq:eom} of the form $u= e^{i(\boldsymbol{k}\cdot\boldsymbol{r}- \omega t)}$, where $\boldsymbol{r}$ is the position vector, and $\boldsymbol{k}$ and $\omega$ are, respectively, the phonon wavevector and angular frequency, yields the secular equation

\begin{equation}
|\Gamma_{ik}-\rho v^2 \delta_{ik}| = 0,
\label{eq:Christoffel}
\end{equation}
where $\Gamma_{ik} = C_{ijkl} n_{j} n_{l}$ is the Christoffel matrix, $n_j$ and $n_l$ are direction cosines, and $\delta_{ik}$ is the Kronecker delta  \cite{haye2012}.

For monoclinic symmetry the elastic constants tensor takes the form \cite{Nye},

\begin{equation}
C_{ij} = \begin{bmatrix}
 C_{11} & C_{12} & C_{13} & 0 & C_{15} & 0 \\
 C_{12} & C_{22} & C_{23} &0 & C_{25} & 0 \\
 C_ {13} & C_{23} & C_{33} & 0 & C_{35} & 0 \\
  0 & 0 & 0 & C_{44} & 0 & C_{46}\\
  C_{15} & C_{25} & C_{35} & 0 & C_{55} & 0\\
  0 & 0 & 0 & C_{46} & 0 & C_{66}\\
\end{bmatrix}.
\label{eq:Monotensor}
\end{equation}
\noindent 
where here the standard monoclinic orientation with $ a \neq b \neq c $ with the unique $b$-axis and $\beta \neq 90^\circ$ used. Figure \ref{fig:bioopt} shows the orientation for a monoclinic biotite structure. Voigt notation has been introduced to reduce the number of subscripts on the elastic constants from four to two by replacing each of the couples $ij$ and $kl$ with a single subscript via the scheme: 11{ $\rightarrow$ 1}; 22{ $\rightarrow$ 2}; 33{$\rightarrow$ 3}; 23, 32{$\rightarrow$ 4}; 13, 31{ $\rightarrow$ 5}; and 12, 21{ $\rightarrow$ 6} \cite{malgrange2014symmetry}. 

For elastic waves propagating in the $ac$ plane (010) at angle $\theta$ to the crystallographic $c$-axis, $n_x=\sin\theta$, $n_y=0$,  and $n_z=\cos\theta$, and Eq. \ref{eq:Christoffel} becomes for the pure transverse (T) mode,
\begin{equation}
C_{eff}= C_{66}\sin^2\theta + C_{44}\cos^2\theta + 2 C_{46}\sin\theta\cos\theta,
\label{eq:puretransverse}
\end{equation}
and 
\begin{equation}
C_{eff}^{\pm}= \frac{1}{2}(-b \pm \sqrt{b^2-4c}),
\label{eq:CeffEQsolve}
\end{equation}
where,
\begin{equation}
b= - [\sin^2{\theta}(C_{11}+C_{55})+2 \sin{\theta}\cos{\theta}(C_{15}+C_{35})
+\cos^2{\theta(C_{33}+C_{55})}], \nonumber
\end{equation}
and 
\begin{multline*}
c=\sin^4{\theta}(C_{11}C_{55}-C^{2}_{15})+2\sin^{3}{\theta}\cos{\theta}(C_{11}C_{35}-C_{13}C_{15})\\+ 
\sin^{2}{\theta}\cos^{2}{\theta}(C_{11}C_{33}-C^{2}_{13}+2C_{15}C_{35}-2C_{13}C_{55})\\+
2\sin{\theta}\cos^{3}{\theta}(C_{33}C_{15}-C_{13}C_{35})+\cos^{4}{\theta}(C_{33}C_{55}-C^{2}_{35}),\\
\end{multline*}
with effective elastic moduli $C_{eff} = \rho v^2$.  The ``+" term in Eq. \ref{eq:CeffEQsolve} refers to the quasi-longitudinal (QL) mode and the ``-" term refers to the quasi-transverse (QT) mode.

For propagation in the $bc$ crystallographic plane (100), the effective moduli are roots of a cubic equation having the form
\begin{equation} \label{eq: bcplaneeqn}
C_{eff}^3  - \alpha C_{eff}^2 + \beta C_{eff}  - \gamma = 0,
\end{equation}
where
\begin{equation*}
\alpha =  [ (C_{66} + C_{22}) \sin^2\theta + (C_{55}+C_{33})\cos^2\theta +C_{44} ],
\end{equation*}
\begin{multline*}
\beta=\sin^4 \theta (C_{44}C_{66} + C_{22}C_{44} + C_{22}C_{66} - C_{46}^2 )\\ +
\sin^2 \theta \cos^2 \theta (C_{44}C_{66} - C_{23}^2 - C_{25}^2  - 2 C_{23}C_{44}- 2C_{46}(C_{25} + C_{35})\\ +
C_{44}C_{55} + C_{33}C_{66} + C_{22}(C_{33}+C_{55}) - C_{46}^2)\\ + 
\cos^4 \theta (C_{33} C_{44} - C_{35}^2 +C_{33}C_{55} + C_{44}C_{55}),\\
\end{multline*}
and 
\begin{multline*} 
\gamma = [\sin^6 \theta ( C_{22}C_{44}C_{66} - C_{22}C_{46}^2 ) \\+ 
\sin^4 \theta \cos^2 \theta ( 2 C_{22} C_{35}C_{46} + C_{22}C_{44}C_{55} \\ +
  C_{22}C_{33}C_{66} - C_{25}^2 C_{44} - C_{23}^2 C_{66} 
 + 2 C_{25}C_{23}C_{46} + 2 C_{23}C_{46}^2 - 2 C_{44}C_{23}C_{66}) \\  + 
 \cos^4 \theta \sin^2 \theta ( C_{33}C_{55}C_{22} - C_{35}^2 C_{22} - C_{33}C_{46}
  + C_{33}C_{44}C_{66}\\ - C_{25}^2C_{33}  - C_{23}^2 C_{55} +
  2 C_{25}C_{35}C_{44} - 2 C_{25} C_{33}C_{46}\\
 + 2 C_{25}C_{23}C_{35} + 2 C_{23}C_{35}C_{46} - 2 C_{23}C_{44}C_{55} ) \\ + 
 \cos^6 \theta (C_{44}C_{55}C_{33} - C_{35}^2 C_{44}) ].\\
 \end{multline*}
The roots of Eq. \ref{eq: bcplaneeqn} are known from the theory of algebraic equations \cite{mens1997} and are given by
\begin{equation*}
M^{i} = X^{i} + \frac{\alpha}{3},  \hspace{5mm} i = 1,2,3 
\end{equation*}
where 
\begin{equation*}
X^{i} = 2 \sqrt[3]{R} \cos( {\frac{\phi}{3} + (i-1)\frac{2 \pi}{3}}) , \hspace{5mm} i =1,2,3,
\end{equation*}
with 
\begin{equation*}
R = \sqrt{(\frac{1}{3} (\frac{\alpha^2}{3} - \beta  ))^3},
\end{equation*}
and
\begin{equation*}
\phi = \arccos \Big ( \frac{{\frac{2}{27} \alpha^3 - \frac{1}{3}\alpha \beta + \gamma  }}{2 R} \Big ).
\end{equation*}

\section{\label{sec:Experiment}Experimental Details}
\subsection{Samples}

\subsubsection{General Physical Characteristics}
The biotite samples used in this study were natural crystals deliberately selected to have a range of colours to maximize the likelihood that they possessed substantially different levels of major cations Fe and Mg.  Platelets with areas of $\sim1$ cm$^{2}$ and thicknesses of a few hundred $\mu$m were cleaved from these larger bulk samples to allow loading onto the sample stage and to expose a pristine surface for Brillouin light scattering experiments.  For the purposes of comparison, two muscovite samples were also analyzed.

\subsubsection{Chemical Composition}
Chemical analysis was carried out on fragments mounted in epoxy cement with the \{001\} cleavage plane orientated vertically and polished to a final grade of 0.25 micron using diamond abrasive.  The samples were carbon coated and examined initially using a JEOL JSM-7100F Scanning Electron Microscope (SEM).  Reconnaissance chemical analysis by energy dispersive x-ray spectroscopy (Thermo\texttrademark) allowed us to confirm the identification of the mica species and identify potential sites for analysis by electron microprobe.

Following SEM investigation the samples were analysed using a JEOL
JXA-8230 Electron Probe Microalalyser (EPMA) operating at an
accelerating voltage of 15kV and beam current of 20nA in wavelength
dispersive (WDS) mode.  The $K_\alpha$ emission lines were counted
using a TAP diffracting crystal for Si, Al, Mg, and Na, a LIFH crystal for Fe, Ti and Mn, a PETL crystal for K, Ca, and Cl and a LDE1 crystal for F.  Counts were collected for the emission peak and for positions on both the low and high wavelength side of the peak.  Total background counting time was the same as that for the emission peak.  The following standards were used: Si, Na albite; Al, pyrope; K K-feldspar; Ca, Mg diopside; Fe almandine; Ti rutile; Mn rhodenite; Cl tugtupite; F apatite.  A secondary standard (Astimex biotite) was analysed intermittently to test for instrument drift.

Estimated detection limits (3$\sigma$, weight percent) for those elements expected to be present in trace to minor quantities are as follows: Mn, Ti 0.015; Na 0.02, Ca 0.006; Cl 0.01; F 0.07. Table \ref{tab:EDX} presents the mean of the analyses for each sample as weight percent oxide recalculated to 22 oxygen atoms and assigned to the tetrahedral (T), octahedral (O) or large cation (X) sites.  The number of analyses per sample was as
follows: Muscovite $\#$2, one, Biotite $\#$1, six, and ten analyses for each of
Biotite $\#$3, $\#$2 and Muscovite $\#$1.

\begin{table}
  \caption{Chemical composition (weight percent oxides) and calculated
    formulae (atoms per formula unit, apfu) of muscovite and biotite samples determined using electron probe
    microanalysis.  Overall atomic percentages of Fe and Mg are presented in the last two rows and are used throughout the paper to identify the samples.}
  \centering \footnotesize
\begin{tabular}{lccccc} \hline\hline
Analyte & Biotite \#1 & Biotite \#2 & Biotite \#3 & Muscovite \#1 & Muscovite \#2 \\
\hline
SiO$_2$	&	31.1855	&	35.0984	&	39.7346	&	48.8039	&	47.3331 \\
TiO$_2$	&	0.10572	&	2.41863	&	1.16121	&	0.64852	&	0.2562 \\
Al$_2$O$_3$	&	17.405	&	14.7722	&	16.8896	&	31.2563	&	37.382 \\
FeO	&	36.612	&	24.8768	&	3.94769	&	7.0886	&	1.7884 \\
MnO	&	0.60836	&	0.31048	&	0.04437	&	0.23769	&	0.0356 \\
MgO	&	0.34392	&	7.65661	&	23.9435	&	1.04965	&	0.8496 \\
CaO	&	0.2015	&	0.17557	&	0.02239	&	0.0457	&	0.0448 \\
Na$_2$O	&	0.41982	&	0.36057	&	0.15724	&	0.43438	&	0.6551 \\
K$_2$O	&	7.27324	&	8.89984	&	10.0847	&	8.5955	&	9.1772 \\
Cl	&	0.17004	&	0.09353	&	0.02054	&	0.02558	&	0.0241 \\
F	&	1.73391	&	1.3567	&	2.13393	&	0.39759	&	0.179 \\
Total	&	95.2906	&	95.4269	&	97.2366	&	98.4102	&	97.6443 \\ \hline
Ion	&	apfu	&	apfu	&	apfu	&	apfu	&	apfu \\ \hline
Si	&	5.16	&	5.48	&	5.46	&	6.39	&	6.10 \\
Al	&	2.84	&	2.52	&	2.54	&	1.61	&	1.90 \\
Al	&	0.55	&	0.2	&	0.19	&	3.21	&	3.77 \\
Ti	&	0.01	&	0.28	&	0.12	&	0.06	&	0.02 \\
Fe	&	5.06	&	3.25	&	0.45	&	0.78	&	0.19 \\
Mn	&	0.09	&	0.04	&	0.01	&	0.03	&	0.00 \\
Mg	&	0.08	&	1.78	&	4.9	&	0.20	&	0.16 \\
Ca	&	0.04	&	0.03	&	0.00	&	0.01	&	0.01 \\
Na	&	0.13	&	0.11	&	0.04	&	0.11	&	0.16 \\ 
K	&	1.53	&	1.77	&	1.77	&	1.43	&	1.51 \\
Cl	&	0.05	&	0.02	&	0.00	&	0.01	&	0.01 \\
F	&	0.91	&	0.67	&	0.93	&	0.16	&	0.07 \\ \hline
Site	&	Total	&	Total	&	Total	&	Total	&	Total \\ \hline
T	&	8.00	&	8.00	&	8.00	&	8.00	&	8.00 \\
O	&	5.79	&	5.56	&	5.68	&	4.28	&	4.16 \\
X	&	1.70	&	1.91	&	1.81	&	1.55	&	1.68 \\
OH	&	0.95	&	0.69	&	0.93	&	0.17	&	0.08 \\
Total	&	16.44	&	16.16	&	16.42	&	14.00	&	13.92 \\
Fe	At\% (Cations)	&	30.78	&	20.11	&	2.74	& 5.57 &	1.36 \\
\hline 
\bf{Fe At\% (Overall)} & \bf{12.7} & \bf{8.1} & \bf{1.1} & \bf{2.1} & \bf{0.5} \\
\bf{Mg At\% (Overall)} & \bf{0.2} & \bf{4.5} & \bf{12.3} & \bf{0.5} & \bf{0.4} \\
\hline \hline
\end{tabular}
\label{tab:EDX}
\end{table}

\begin{table}
\caption{Principal refractive indices of biotites and muscovite determined using the Becke line test.}
\centering \footnotesize
\begin{tabular}{cccccc c}
\hline
Sample  &  Colour    &  $n_{\alpha}$  &  $n_{\beta}$  &  $n_{\gamma}$  & $n_{avg}$ & $2V$ \\ [0.5ex] \hline
Biotite & Green & 1.551 & 1.665 & 1.665 & 1.613 & $\ll20^{\circ}$ \\
12.7\% Fe, 0.2\% Mg & & & & & & \\ \hline
Biotite & Green & 1.575 & 1.657 & 1.657 & 1.629 & $\ll10^{\circ}$ \\ 
8.1\% Fe, 4.5\% Mg & & & & & & \\ \hline 
Biotite & Pale Brown & 1.544 & 1.597 & 1.597 & 1.579 &Too small \\
1.1\% Fe, 12.3\% Mg & & & & &  &to measure \\ \hline
Muscovite & Very Pale & 1.543 & 1.567 & 1.618 & 1.576& $\sim 35^{\circ}-40^{\circ}$ \\
2.1\% Fe, 0.5\% Mg & Brown-Green & & & & & \\ \hline
Muscovite & Pale Brown & 1.558 & 1.603 & 1.610 & 1.590 &-- \\
0.5\% Fe 0.4\% Mg & & & & & & \\ \hline
Biotite (Ref. \cite{coat1944}) & Green-Grey &1.570 & 1.640&1.640 & 1.620 & -- \\ \hline
Biotite (Ref. \cite{bilg1956}) & -- &1.575 & 1.617 &1.621 & 1.604 & 30$^\circ$\\ \hline
Biotite (Ref. \cite{hutt1947}) & Dark Brown &1.57 & 1.64&1.64 & 1.620 & -- \\ \hline
Biotite (Ref. \cite{lars1938}) & -- &1.586 & 1.643 &1.643 & 1.623 & 0$^\circ$ - 8$^\circ$ \\ \hline
Biotite (Ref. \cite{nock1939}) & Blue-Green &1.582 & 1.625&1.625 & 1.610 & -- \\ \hline
Biotite (Ref. \cite{pagl1940}) & Violet-Brown &1.544 & 1.583&1.585 & 1.571 & -- \\

\hline
\end{tabular}
\label{tab:EDS1}
\end{table}
\subsubsection{Refractive Indices}
\begin{figure}
  \centering
  \includegraphics[scale = 0.5]{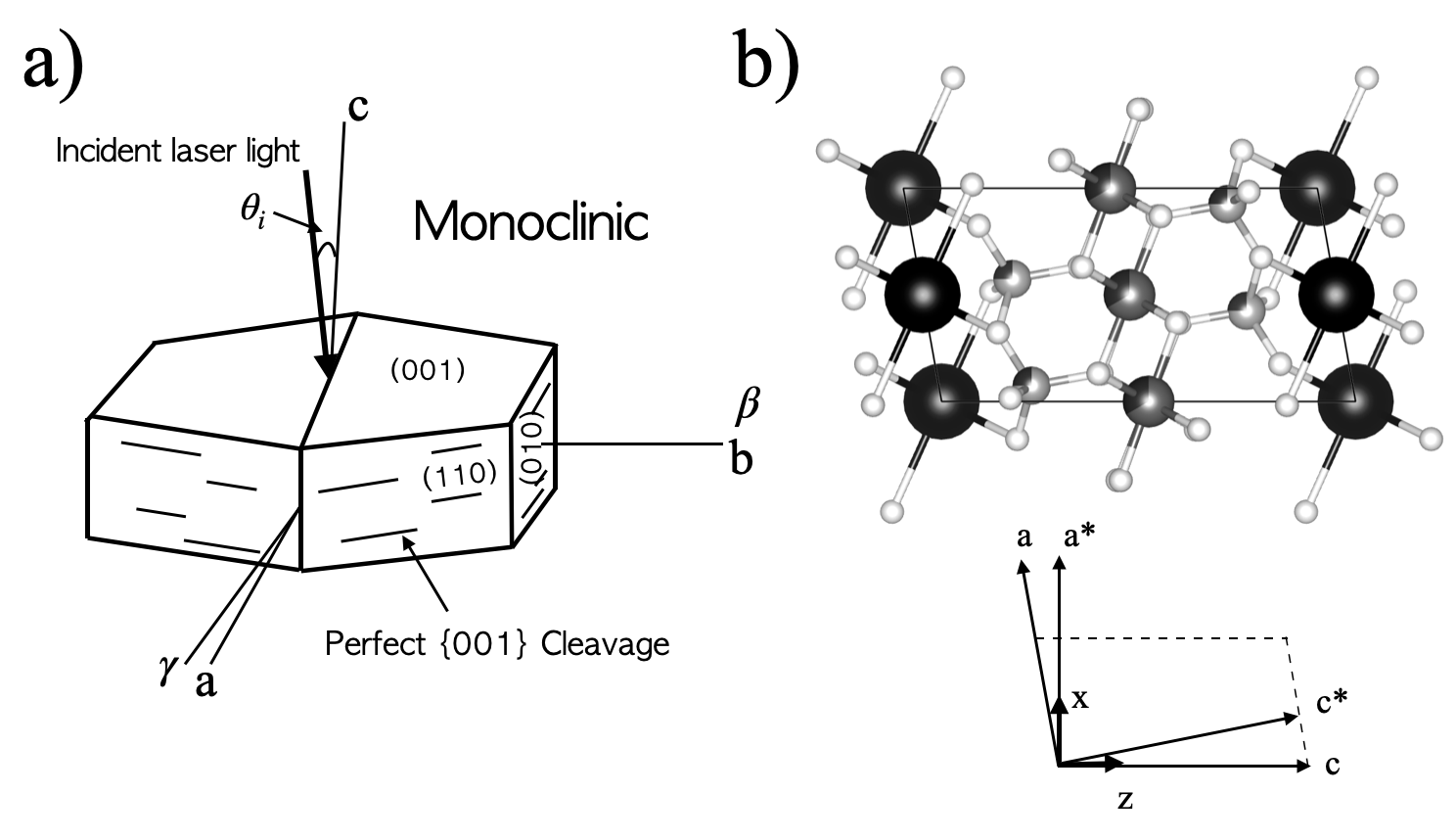}
  \caption{ \textbf{a}) $a$, $b$ and $c$ are the crystallographic axes, $\beta$ and $\gamma$ are the principal refractive indices. Incident laser beam shown with $\theta_i$ indicating angle from $c$-axis. Not shown principle refractive index $\alpha$.
  \textbf{b)} Biotite structure (unique $b$ axis and $\beta \neq 90^\circ$) \cite{takeda1975mica} viewed down from \textit{b}. Projection of unit cell shown along with the Cartesian coordinate system, z $\parallel$ $c$, x $\parallel$ ${a^*}$ and y $\parallel$ $b$. Note: $b \parallel b^*$. }
  \label{fig:bioopt}
\end{figure}

Fig. \ref{fig:bioopt} shows the optical orientation of biotite, with
the crystallographic {\it b} axis chosen coincident with the two fold
symmetry axis.  Refractive indices of the biotite samples were
obtained using the Becké line test \cite{ness2012}. Polarized light
was provided by a Nikon Eclipse i50 Pol polarizing light microscope
and Cargille optical liquids of known refractive indices and graduated
at intervals of 0.002 were used as immersion liquids.  Measurements of
refractive indices $\gamma$ and $\beta$ were made on sample fragments
that were placed on the perfect \{001\} cleavage.  The remaining refractive
index $\alpha$ was determined by mounting each sample on a spindle
stage with cleavage plane perpendicular to the stage
surface. Refractive index values determined using this approach are
presented in Table \ref{tab:EDS1} and are considered accurate to $\pm$
0.001.

\subsubsection{Density Determination}
\begin{figure}
  \centering
  \includegraphics[scale = 0.4]{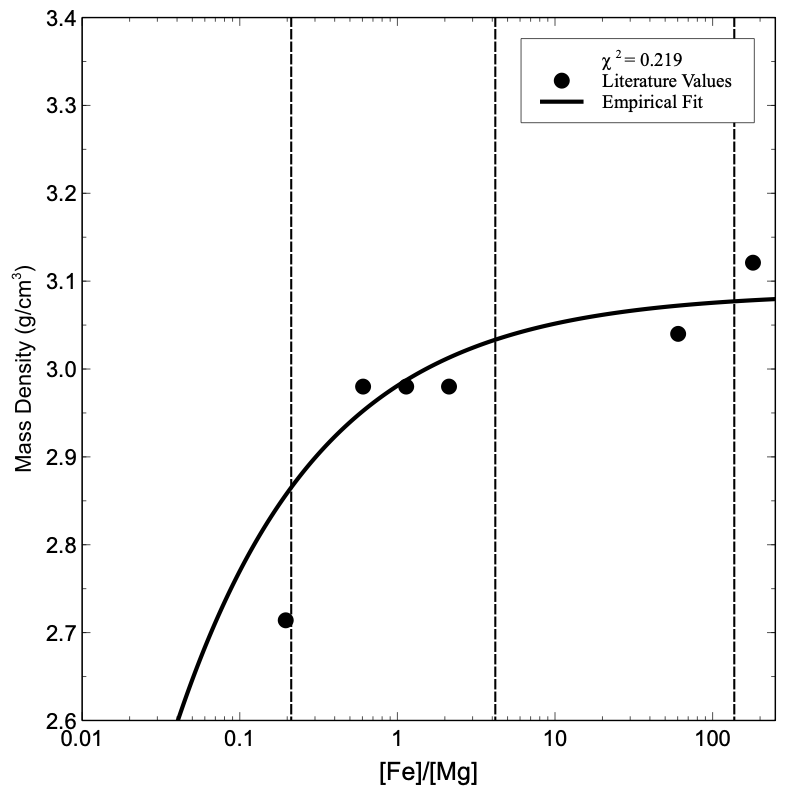}
  \caption{Mass density versus $x =$ [Fe]/[Mg] data for biotites of known composition \cite{coat1944,bilg1956,hutt1947,lars1938,nock1939,pagl1940}. The dashed vertical lines indicate [Fe]/[Mg] ratios for the biotites of the present work.}
  \label{fig:density}
\end{figure}

Mass densities for the biotite samples used in this work were required to determine elastic constants.  The equation $\rho = Ax^{B} + C$, with $A=-0.141$, $B=-0.523$, and $C=3.085$, was found to yield a good empirical fit to published density versus $x =$ [Fe]/[Mg] data for biotites of known composition \cite{coat1944,bilg1956,hutt1947,lars1938,nock1939,pagl1940}. This was used, together with the composition data obtained from electron probe microanalysis, to estimate the density for the samples of this work. Figure \ref{fig:density} shows the fitting results with $\displaystyle{\chi^2}$ quoted in the legend.  The uncertainty in density determined using this approach is $\sim5$\%.  It is important to note that the empirical equation above should not be used to estimate the density of a biotite for which [Fe]/[Mg] lies outside the range of those of the samples studied in present work as it may yield inaccurate or unphysical values, particularly for small [Fe]/[Mg] values.  

The density of muscovite was taken as fixed and set equal to the average of two previously published values: $(2832 + 2844)/2$ kg$\cdot$m$^{-3}$ = 2838 kg$\cdot$m$^{-3}$ \cite{vaug1986,mcne1993}.

\subsection{Brillouin Light Scattering Apparatus}
Brillouin spectra were obtained under ambient conditions utilizing a $180^\circ$ backscattering geometry.  Light of wavelength $\lambda_{i} = 532$ nm and power of 60 mW from a Nd:YV$O_4$ single mode laser was incident on the target sample at angles $5^{\circ} \lesssim \theta_{i} \lesssim 75^{\circ}$, corresponding to probed phonon propagation directions ranging from $\sim3^{\circ}$ to $\sim37^{\circ}$ from the crystallographic $c$-axis.  Focusing of incident light onto the sample and collection of scattered light was accomplished using a high-quality anti-reflection-coated camera lens of focal length $f=5$ cm and $f/\# = 5.6$.  After exiting this lens, the scattered light was processed by a spatial filter (40 cm lens - 450 $\mu$m-diameter pinhole - 20 cm lens) and subsequently frequency-analyzed by an actively-stabilized 3+3 pass tandem Fabry-Perot interferometer (JRS Scientific Instruments).  The free spectral range of the interferometer was set to 40 GHz and the finesse was $\sim100$. The light transmitted by the interferometer was incident on a pinhole of diameter 700 um and detected by a low-dark count rate ($\lesssim 1$ s$^{-1}$)  photomultiplier tube where it was converted to an electrical signal and sent to a computer for storage and display. A schematic diagram of the apparatus can be found in Ref. \cite{andr2018}.

\section{\label{sec:Results}Results \& Discussion}

\subsection{Spectra}
\begin{figure}
\centering
\includegraphics[scale = 0.5]{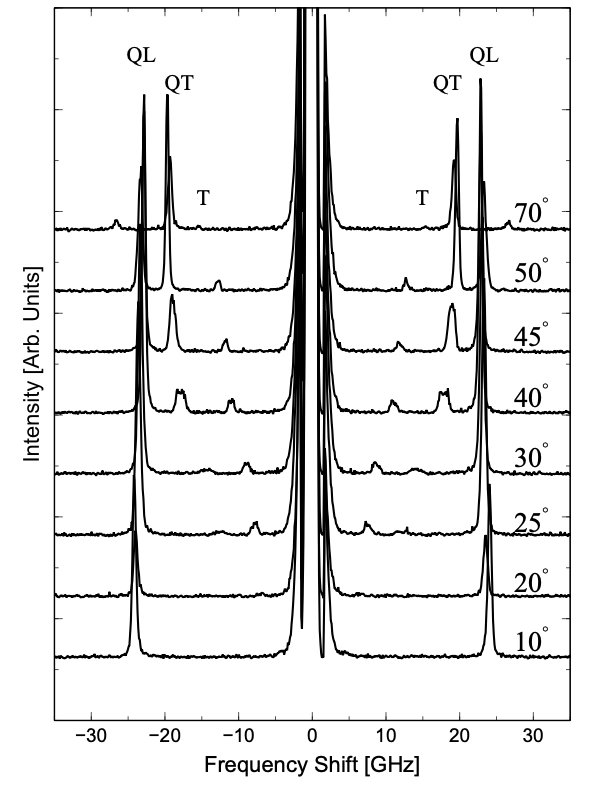}
  \caption{Brillouin spectra of an Fe-rich biotite crystal (12.7\% Fe, 0.2\% Mg).  The probed phonons propagated in the (010) plane.  Angles of incidence are indicated. T, QT, and QL refer to peaks due to pure transverse, quasi-transverse, and quasi-longitudinal acoustic modes, respectively.}
  \label{fig:figure4.1}
\end{figure}

Figure \ref{fig:figure4.1} shows a series of Brillouin spectra collected from the Fe-rich biotite crystal containing 12.7\% Fe and 0.2\% Mg.   Three sets of Brillouin doublets were observed and attributed to T, QT, and QL acoustic modes due to the similarity of the frequency shifts to those of muscovite \cite{vaug1986,mcne1993}.  The frequency shifts of all of these peaks showed significant variation with angle of incidence, reflecting the expected large elastic anisotropy of biotite. Spectra of other samples were qualitatively similar, although the T peaks in the sample with 8.1\% Fe and 4.5\% Mg were noticeably weaker than for the other samples.

\subsection{Elastic Wave Velocities}
\begin{figure}[H]
\centering
\includegraphics[scale=0.32]{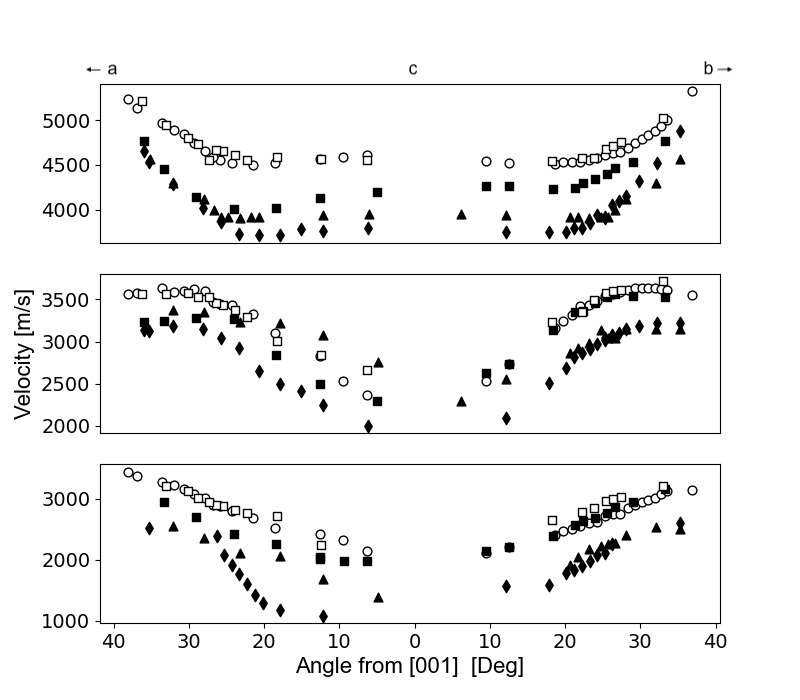}
  \caption{Quasi-longitudinal (top), quasi-transverse (center), and pure transverse (bottom) mode velocities versus direction ({\it i.e.}, angle from the crystallographic $c$-axis) in the $ac$ and $bc$ crystallographic planes. $\circ$ -  muscovite 2.1\% Fe, 0.5\% Mg; $\square$ - muscovite 0.5\% Fe, 0.4\% Mg; $\blacksquare$ - biotite 1.1\% Fe, 12.3\% Mg; $\blacktriangle$ - biotite 8.1\% Fe, 4.5\% Mg; $\blacklozenge$ -  biotite 12.7\% Fe, 0.2\% Mg.  The vertical and horizontal error bars are approximately the size of the symbols.}
\label{velvsdir}
\end{figure}

Figure \ref{velvsdir} shows the T, QT, and QL mode velocities in the $ac$ and $bc$ crystallographic planes as measured from the $c$-axis.  These velocities were calculated from Eq. \ref{brillouineqn} using the associated Brillouin peak frequency shifts and, because birefringence effects were negligible (no obvious peak splitting), the average refractive index for each sample \cite{ness2012}.   
For propagation directions near the $c$-axis, the T and QT velocities are relatively low and increase with increasing angle away from the $c$-axis in both the $ac$ and $bc$ planes. The QL mode velocity remains relatively constant for angles close to the $c$-axis and increases for propagation directions greater than $\approx 20-25^\circ$. Studies on muscovite  \cite{vaug1986} and phlogopite \cite{chhe2014} show a similar dependence of sound velocities on propagation direction over common ranges.

It can also be seen from Figure \ref{velvsdir} that the elastic wave velocities for biotite, nearly without exception, increase with decreasing Fe concentration (or, equivalently, increasing Mg concentration), and approach those of muscovite for the biotite sample with the lowest Fe concentration. The same type of behaviour was observed in ultrasonic studies on muscovite, phlogopite, and a biotite of unknown composition, where, for nearly all propagation directions for which measurements were made, the velocities for the phlogopites and the biotite were lower than the corresponding velocities for muscovite \cite{alek1961}. This may be due in part to the relatively large atomic mass of Fe compared to Mg and Al, resulting in a lower vibrational frequency and therefore a lower velocity, but it could also be due to other factors such as differences in chemical bonding. 

\begin{landscape}

\begin{table}
\scriptsize
\vspace{30mm}

\caption{Elastic constants (in GPa) and selected elastic constant ratios for biotite and muscovite obtained in the present work and for biotite, muscovite, and phlogopite obtained in previous studies.}

\begin{tabular}{c  c  c  c  c  c  c  c  c  c  c  c c c c c c}
\hline
Sample & $C_{11}$ &$C_{22}$ & $C_{33}$ & $C_{44}$ & $C_{55}$ & $C_{66}$ & $C_{12}$ & $C_{13}$ & $C_{15}^{a}$ & $C_{23}$ & $C_{25}^{a}$ & $C_{35}^{a}$ & $C_{46}^{a}$ & $C_{22}/C_{11}^{b}$ & $C_{55}/C_{44}^{b}$ & $C_{23}/C_{13}^{b}$ \\
 \hline
Biotite  & 178 & 176  & 42 & 4.4 & 11 & 77 & 24 & 14 & -9.3 & 22 & 27 & 1.8 & -7.7 & 0.99 & 2.50 & 1.57 \\ 
12.7\% Fe, 0.2\% Mg & $\pm$ 5.4 & $\pm$ 3.6 & $\pm$ 0.3 & $\pm$ 0.2 & $\pm$ 0.8 & $\pm$ 1.3 & - & $\pm$ 6.9 & $\pm$ 4.6 & $\pm$ 5.2 & $\pm$ 1.5& $\pm$ 0.9& $\pm$ 3.8 & $\pm$ 0.05 & $\pm$ 0.30 & $\pm$ 1.14 \\ \hline

Biotite  & 183 & 183 & 48 & 7.9 & 16 & 75 & 33 & 21 & -12 & 22 & 27 & -2.7 & -6.2 & 1.00 & 2.02 & 1.04 \\ 
8.1\% Fe, 4.5\% Mg & $\pm$ 5.4 & $\pm$ 5.5 & $\pm$ 0.6 & $\pm$ 0.2 & $\pm$ 0.8 & $\pm$ 1.7 & - & $\pm$ 4.7 & $\pm$ 2.1 & $\pm$ 4.5 & $\pm$ 3.5& $\pm$ 1.3& $\pm$ 2.2 & $\pm$ 0.06 & $\pm$ 0.15 & $\pm$ 0.44  \\ \hline
 
Biotite  & 178 & 171 & 52 & 12.2 & 17 & 76 & 26 & 28 & -13 & 22 & 22 & -8.5 & -6.2 & 0.96 & 1.39 & 0.78 \\
1.1\% Fe, 12.3\% Mg & $\pm$ 5.2 & $\pm$ 3.5 & $\pm$ 1.2 & $\pm$ 0.3 & $\pm$ 0.5 & $\pm$ 1.5 & - & $\pm$ 5.4 & $\pm$ 1.6 & $\pm$ 5.1 & $\pm$ 2.5& $\pm$ 1.9& $\pm$ 3.4& $\pm$ 0.05 & $\pm$ 0.08 & $\pm$ 0.33  \\ \hline

Muscovite & 182 & 182 & 66 & 13.8 & 17 & 71 & 40 & 22 & -2.3 & 24 & 18 & -3.8 & 1.3 & 1.00 & 1.23 & 1.09 \\ 
2.1\% Fe, 0.5\% Mg & $\pm$ 4.0 & $\pm$ 3.8 & $\pm$ 1.1 & $\pm$ 0.5 & $\pm$ 0.6 & $\pm$ 1.4 & - & $\pm$ 2.2 & $\pm$ 3.3 & $\pm$ 2.7 & $\pm$ 3.4& $\pm$ 1.7& $\pm$ 2.0 & $\pm$ 0.04 & $\pm$ 0.09 & $\pm$ 0.23  \\ \hline

Muscovite &  180 & 183 & 56 & 17.3 & 23 & 77 & 26 & 22 & -12 & 19 & 29 & -2.7 & -7.2 & 1.02 & 1.33 & 0.86 \\
0.5\% Fe, 0.4\% Mg & $\pm$ 3.6 & $\pm$ 3.6 & $\pm$ 1.2 & $\pm$ 0.5 & $\pm$ 0.6 & $\pm$ 1.3 & - & $\pm$ 1.9 & $\pm$ 2.9 & $\pm$ 4.6 & $\pm$ 4.5& $\pm$ 1.0 & $\pm$ 2.2 & $\pm$ 0.04 & $\pm$ 0.07 & $\pm$ 0.28  \\ \hline

Muscovite (Composition Unknown) & 181.0 & 178.4 & 58.6 & 16.5 & 19.5 & 72.0 & 48.8 & 25.6 & -14.2 & 21.2 & 1.1 & 1.0 & -5.2 & 0.99 & 1.18 & 0.83\\
\cite{vaug1986} & $\pm$ 1.2 & $\pm$ 1.3 & $\pm$ 0.6 & $\pm$ 0.6 & $\pm$ 0.5 & $\pm$ 0.7 & $\pm$ 2.5 & $\pm$ 1.5 & $\pm$ 0.8 & $\pm$ 1.8 & $\pm$ 3.7 & $\pm$ 0.6 & $\pm$ 0.9 & $\pm$ 3.7 & $\pm$ $0.6^{c}$ & $\pm$ $0.9^{c}$  \\ \hline

Muscovite (Composition Unknown) & 176.5 & 179.5 & 60.9 & 15.0 & 13.1 & 70.7 & 47.7 & 20.0 & -1.2 & 23.0 & 11.1 & -0.7 & 0.7 & 1.02 & 0.87 & 1.15 \\
\cite{mcne1993} & $\pm$ 1.1 & $\pm$ 1.3 & $\pm$ 0.6 & $\pm$ 0.3 & $\pm$ 0.2 & $\pm$ 0.6 & $\pm$ 1.2 & $\pm$ 1.1 & $\pm$ 0.6 & $\pm$ 8 &$\pm$ 5.3 & $\pm$ 0.5 & $\pm$ 0.5 & $\pm$ 0.01 & $\pm$ $0.07^{c}$ & $\pm$ $0.12^{c}$ \\ \hline

Phlogopite (DFT-GGA) & 181.2 & 184.7 & 62.1 & 13.5 & 20.0 & 67.9 & 47.6 & 12.2 &-15.7 & 12.1 & -4.9 & -1.2 & -5.9 & 1.02 & 1.48 & 0.99 \\
\cite{chhe2014} \\ \hline

Muscovite (DFT-GGA) & 180.9 & 170.0 & 60.3 & 18.4 & 23.8 & 70.5 & 53.4 & 27.4 &-14.7 & 23.5 & 1.4 & -1.0 & -1.8 & 1.06 & 1.29 & 0.86 \\
\cite{militzer2011first} \\ \hline

Phlogopite (DFT-LDA) & 199.5 & 201.2 & 82.2 & 17.0 & 25.3 & 72.4 & 54.1 & 25.4 &-13.1 & 24.4 & -4.5 & -2.8 & -6.4 & 1.01 & 1.49 & 0.96 \\
\cite{chhe2014} \\ \hline

Muscovite (DFT-LDA) & 194.3 & 188.0 & 91.1 & 25.2 & 30.5 & 71.3 & 68.1 & 43.2 &-14.3 & 39.5 & 1.7 & 1.1 & -0.6 & 1.03 & 1.21 & 0.91 \\
\cite{militzer2011first} \\

\hline
\multicolumn{17}{l}{$^{a}$Elastic constant is zero for hexagonal symmetry.}\\
\multicolumn{17}{l}{$^{b}$Ratio is unity for hexagonal symmetry.}\\ \multicolumn{17}{l}{$^{c}$Uncertainties in ratios calculated by the authors of the present work using standard error formulae.}
\end{tabular}
\label{tab:elasticconstants}
\end{table}
\end{landscape}

\begin{table}[t]
\centering
\small
\caption{Voigt and Reuss bulk ($K_X$) and shear moduli ($G_X$) (in GPa) for biotite and muscovite obtained in the present work and for biotite, muscovite, and phlogopite obtained in previous studies.}
\begin{tabular}{l c  c  c  c} \hline \hline
Sample & K$_V$ & K$_R$ & G$_V$ & G$_R$\\ \hline
Biotite - 12.7\% Fe, 0.2\% Mg & 57.3 $\pm$ 4.2 & 35.3 $\pm$ 2.8  & 41.0 $\pm$ 3.0 & 9.44 $\pm$ 0.8 \\
Biotite - 8.1\% Fe, 4.5\% Mg  & 63.0 $\pm$ 4.3 & 42.0 $\pm$ 2.9 & 43.0 $\pm$ 3.0 & 16.0 $\pm$ 1.1 \\
Biotite - 1.1\% Fe, 12.3\% Mg  & 61.4 $\pm$ 4.1 & 43.0 $\pm$ 2.9 & 42.9 $\pm$ 2.7 & 19.4 $\pm$ 1.2 \\
Muscovite - 2.1\% Fe, 0.5\% Mg & 66.9 $\pm$ 5.0 & 43.3 $\pm$ 3.2 & 52.0 $\pm$ 3.9 & 25.0 $\pm$ 1.9\\
Muscovite - 0.5\% Fe, 0.4\% Mg & 61.0 $\pm$ 4.8 & 45.8 $\pm$ 3.6 & 46.9 $\pm$ 3.7 & 26.9 $\pm$ 2.1\\
Muscovite (Comp Unknown) \cite{vaug1986} & 67.7 $\pm$ 1.5 & 48.7 $\pm$ 1.1 & 43.1 $\pm$ 0.94 & 27.6 $\pm$ 0.6 \\
Muscovite (Comp Unknown) \cite{mcne1993} & 66.5 $\pm$ 2.6 & 49.0 $\pm$ 1.8 & 42.0 $\pm$ 1.6 & 23.0 $\pm$ 0.9  \\
Phlogopite (DFT-GGA)\cite{chhe2014} & 61.0 $\pm$ 4.7 & 43.0 $\pm$ 8.2 & 43.3 $\pm$ 1.3 & 24.1 $\pm$ 2.9\\ 
Muscovite (DFT-GGA) \cite{militzer2011first} & 86.1 & 73.1 & 47.0 & 38.2 \\
Phlogopite (DFT-LDA) \cite{chhe2014} & 77.2 $\pm$ 5.6 & 62.9 $\pm$ 7.2 & 47.6 $\pm$ 1.1 & 31.2 $\pm$ 2.0 \\ 
Muscovite (DFT-LDA) \cite{militzer2011first} & 68.8 & 50.1 & 43.0 & 31.0 \\ \hline \hline
\end{tabular}
\label{tab:bulkshearmoduli}
\end{table}

\subsection{Elastic Constants}
\subsubsection{Determination of Elastic Constants}
\begin{figure}[H]
\centering
\includegraphics[scale = 0.40]{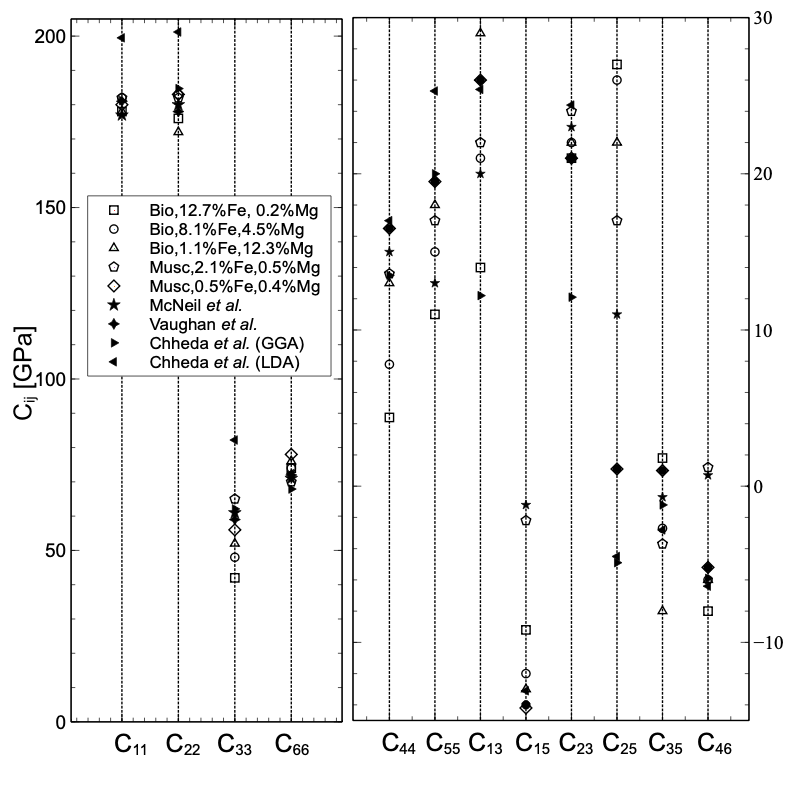}
  \caption{Elastic constants of biotite and muscovite obtained in the present work (open symbols) and previous studies (filled symbols). $\protect\square$ - biotite 12.7\% Fe, 0.2\% Mg; $\protect\medcirc$ - biotite 8.1\% Fe, 4.5\% Mg; $\protect\bigtriangleup$ - biotite 1.1\% Fe, 12.3\% Mg; $\protect\pentagon$ - muscovite 2.1\% Fe, 0.5\% Mg; $\protect\Diamond$ - muscovite 0.5\% Fe, 0.4\% Mg; $\protect\bigstar$ - muscovite \cite{mcne1993}; $\protect\vardiamondsuit$ - muscovite \cite{vaug1986}; $\protect\blacktriangleright$ - phlogopite \cite{chhe2014} DFT-GGA; $\protect\blacktriangleleft$ - phlogopite \cite{chhe2014} DFT-LDA.  Vertical error bars are approximately the size of the symbols for elastic constants of the present work.}
\label{fig:Cefftrends}
\end{figure}

Table \ref{tab:elasticconstants} and Fig.~\ref{fig:Cefftrends} give the elastic constants for the biotite and muscovite samples of the present work along with values obtained in previous studies. All of the constants for the samples of the present work except $C_{12}$ were obtained using a custom non-linear least-squares fitting routine written in MATLAB and based on the Levenberg-Marquardt algorithm. This routine executed a global fit of the expressions for $C_{eff}=\rho v_{m}^{2}(\theta, C_{ij})$, where $m$ = T, QT, and QL, given in Sec. \ref{elasticwaves} to experimental data by minimizing the square of the difference between experimental and calculated $\rho v_{m}^{2}(\theta, C_{ij})$ values for all three modes and measured propagation directions in the $ac$ and $bc$ planes simultaneously through adjustment of the $C_{ij}$.  Initial guesses of elastic constants were those determined for biotite with assumed hexagonal symmetry and muscovite found in the literature \cite{vaug1986,mcne1993,alek1961}.  Embedded in the fitting routine was the constraint that the elastic constants, when appropriately combined, satisfy the conditions for elastic stability ({\it i.e.}, that the elastic energy be positive) for a crystal with monoclinic symmetry \cite{mouhat2014necessary}.  These are that the diagonal elements of the elastic constants tensor, $C_{ii}$, be greater than zero and that  

\begin{multline}
    (C_{11}C_{25}^2C_{33} + C_{15}^2(-C_{23}^2 + C_{22}C_{33})\\ - 2C_{11}C_{23}C_{25}C_{35} - C_{12}^2C_{35}^2 + C_{11}C_{22}C_{35}^2\\ + 2C_{15}(C_{13}C_{23}C_{25} - C_{12}C_{25}C_{33} - C_{13}C_{22}C_{35} \\+ C_{12}C_{23}C_{35}) + C_{11}C_{23}^2C_{55} + C_{12}^2C_{33}C_{55}\\ - C_{11}C_{22}C_{33}C_{55} + C_{13}^2(-C_{25}^2 + C_{22}C_{55}) \\+ 2C_{12}C_{13}(C_{25}C_{35}-C_{23}C_{55}))(C_{46}^2 - C_{44}C_{66}) > 0, 
\label{stabcon1}
\end{multline} 

\begin{multline}
C_{44}(-C_{11}C_{25}^2 C_{33} + C{15}^2(C_{23}^2 - C_{22}C_{33})\\ + 2C_{11}C_{23}C_{25}C_{35} + C_{12}^2C_{35}^2 - C_{11}C_{22}C_{35}^2\\ - 2C_{15}(C_{13}C_{23}C_{25} - C_{12}C_{25}C_{33} - C_{13}C_{22}C_{35}\\ + C_{12}C_{23}C_{35}) - C_{11}C_{23}^2C_{55} - C_{12}^2C_{33}C_{55}\\ + C_{11}C_{22}C_{33}C_{55} + C_{13}^2(C_{25}^2 - C_{22}C_{55})\\ + 2C_{12}C_{13}(-C_{25}C_{35} + C_{23}C_{55})) > 0,
\label{stabcon2}
\end{multline} 

\begin{equation}
-C_{44}(C_{13}^2C_{22} - 2C_{12}C_{13}C_{23} + C_{12}^2C_{33} + C_{11}(C_{23}^2 - C_{22}C_{33})) > 0,
\label{stabcon3}
\end{equation} 

\begin{equation}
C_{11}C_{22}C_{33} - C_{13}^2C_{22} + 2C_{12}C_{13}C_{23} - C_{11}C_{23}^2 - C_{12}^2C_{33}  > 0,
\label{stabcon4}
\end{equation}
and
\begin{equation}
C_{11}C_{22} - C_{12}^2 >0.
\label{stabcon5}
\end{equation}

\begin{figure}
\centering
\includegraphics[scale=0.55]{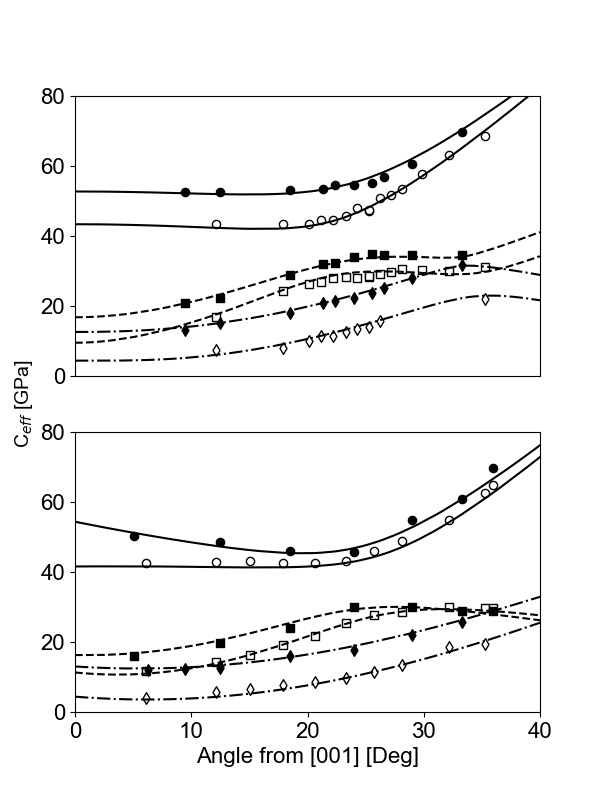}
  \caption{Effective elastic moduli, $C_{eff}=\rho v^{2}$, versus angle from the [001] direction for (i) Unfilled symbols - Fe-rich biotite (12.7\% Fe, 0.2\% Mg), and (ii) Filled symbols - Fe-poor biotite (1.1\% Fe, 12.3\% Mg). Upper plot - $bc$ plane; Lower plot - $ac$ plane. Solid ( -- ), dashed (-- --) and dash-dotted (-- $\cdot$ --) lines are best-fit curves for quasi-longitudinal, quasi-transverse, and pure transverse modes, respectively. $\circ$, $\bullet$ - Quasi-longitudinal mode; $\square$, $\blacksquare$ - Quasi-transverse mode; $\diamondsuit$, $\blacklozenge$ - Pure transverse mode.}
  \label{fig:Ceff_Fit_Fe_Mg}
\end{figure}

\begin{figure}
\centering
\includegraphics[scale=0.49]{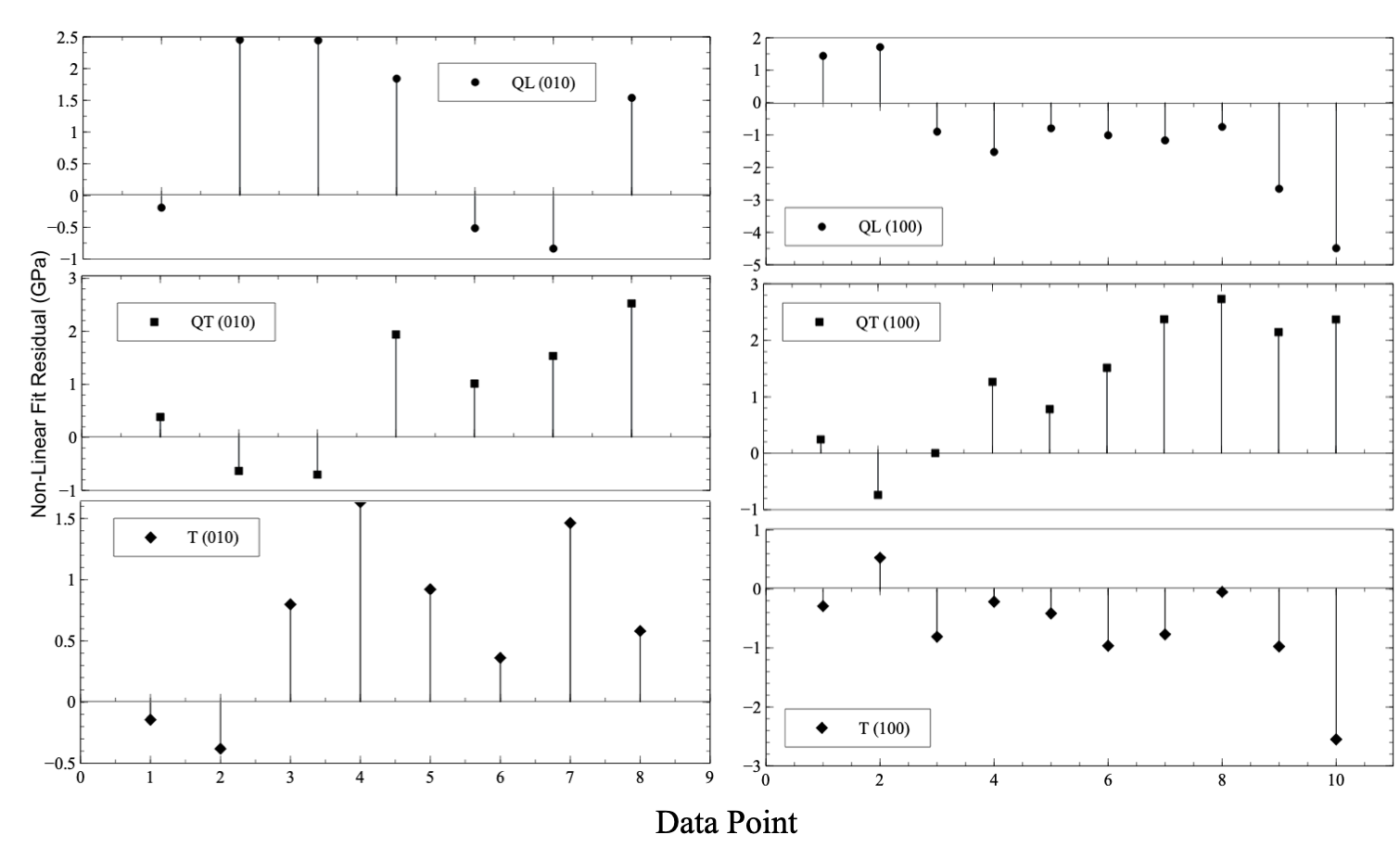}
  \caption{Residuals plots for Fe-poor biotite (1.1\% Fe, 12.3\% Mg).  Left panels - {\it ac} plane.  Right panels - {\it bc} plane.  The standard error of regression is 1.51 GPa.}
  \label{fig:ResidualPlot}
\end{figure}

In all cases, the quality of the fits was excellent as can be seen for Fe-rich and Fe-poor biotite in Fig. \ref{fig:Ceff_Fit_Fe_Mg}. The standard error of regression (SER) for the highest- and lowest-quality least-squares fits have been calculated as points of reference. The best fit (SER = 1.28 GPa) was obtained for Fe-rich biotite (12.7\% Fe, 0.2\% Mg) and the lowest-quality fit (SER = 1.69) was obtained for the biotite sample with 8.1\% Fe and 4.5\% Mg. Fig. \ref{fig:ResidualPlot} shows a representative series of residual plots for the Fe-poor (1.1\% Fe, 12.3\% Mg) sample.  The uncertainties in the elastic constants were estimated by averaging the difference between individual best-fit constants and corresponding members of second and third sets of elastic constants obtained by increasing and decreasing, respectively, all experimental $\rho v_{m}^{2}(\theta, C_{ij})$ values by 2\% and re-executing the fitting routine.

$C_{12}$ could not be determined using the least-squares fitting procedure described above because it does not appear in the expressions for $C_{eff}$ given in Sec.\;\ref{elasticwaves}.  Estimates of this constant, however, were obtained by approximating the symmetry of biotite and muscovite as hexagonal, in which case $\displaystyle C_{12} = C_{11} - 2C_{66}$.  It is emphasized that this expression is used here only as a means to estimate $C_{12}$ and is not expected to hold in general for the samples of the present study due to the true symmetry being monoclinic.

Table \ref{tab:bulkshearmoduli} presents Voigt and Reuss bulk and shear moduli for the biotite and muscovite samples of the present work determined from the elastic constants.  Also included for the purposes of comparison are values of these quantities for muscovite and phlogopite from previously published studies \cite{vaug1986, mcne1993, chhe2014, militzer2011first}. 

\subsubsection{Monoclinic Character}
The non-zero values of $C_{15}$, $C_{25}$, $C_{35}$, and $C_{46}$, along with the fact that $C_{44} \neq C_{55}$, and $C_{13} \neq C_{23}$ (see Table \ref{tab:elasticconstants}) reaffirm the monoclinic character of biotite. $C_{15}$, $C_{25}$, $C_{35}$, and $C_{46}$ all differ significantly from zero with $C_{15}$ and $C_{46}$ having negative values for each of the three biotites and $C_{35}$ being negative for two of the samples.  Interestingly, while $C_{11} \simeq C_{22}$ for each of the biotite samples, $C_{44}$ is considerably smaller than $C_{55}$ for all three, with the ratio $C_{55}/C_{44}$ decreasing from about 2.5 to 1.4 as the Fe content decreases.  Similar behaviour is seen for $C_{23}/C_{13}$, which ranges from $\sim1.6$ for Fe-rich biotite to $\sim0.8$ for biotite with low Fe content.  Collectively, these results indicate a relatively high degree of elastic anisotropy, ostensibly concentrated in the shear properties, and highlight the deficiencies associated with the previously-invoked approximation of hexagonal symmetry \cite{alek1961}.

\subsubsection{Dependence on Primary Cation Concentration}
As can be seen in Table \ref{tab:elasticconstants} and Figure \ref{fig:Cefftrends}, the dependences of biotite elastic constants on primary cation concentration are varied. $C_{11}$, $C_{22}$, $C_{66}$, and $C_{23}$ are nearly independent of Fe (or Mg) concentration, taking on very similar values for all three samples of the present work (ranges: $C_{11}\sim3$\%, $C_{22}\sim7$\%, $C_{66}\sim3$\%, $C_{23}\sim0$\%).  The remaining constants show significant dependence on [Fe] (see Fig. \ref{fig:Cij_vs_Fe_Conc}). $C_{33}$, $C_{44}$, $C_{55}$, and $C_{13}$ all increase with decreasing Fe concentration, while $C_{15}$ and $C_{35}$ decrease with decreasing Fe concentration (see Table \ref{tab:Cij_vs_[Fe]_fits}. $C_{25}$ has the same value for the two biotites of higher Fe concentration, with the value for the sample with the smallest Fe concentration being lower by $\sim15\%$.  In contrast, $C_{46}$ is lowest for the Fe-rich biotite and takes on a common, $\sim20\%$ higher value for the two samples of lower Fe concentration. 

The composition information in Table \ref{tab:EDX} shows that, by far, the most significant sample-to-sample difference among the biotite crystals is in Fe and Mg content, suggesting that the behaviour of the elastic constants is a manifestation of the structural response of the crystal to Fe-Mg substitution in the octahedral sheets, a well-known and common substitution mechanism in biotite. The structural changes that occur as a result of this substitution are functions of cation radius and include rotation of tetrahedra and flattening of octahedra in the tetrahedral and octahedral sheets, respectively \cite{cibi2005,hewi1975,tomb2002}. Elastic constants  $C_{11}$, $C_{22}$, and $C_{66}$ will be relatively insensitive to these changes due to the strong intralayer bonding within a tetrahedral-octahedral-tetrahedral (TOT) sheet. In contrast, the weak interlayer bonding between adjacent TOT sheets will, in general, result in the remaining constants displaying a stronger dependence on Fe concentration.  With the exception of $C_{23}$, this is precisely the behaviour that is observed.

\begin{table}[t]
\centering

\caption{Selected elastic constants ratios versus Fe concentration for biotite samples of the present work. C$_{ij}$ / C$_{ij}^{Fe-Poor}$ = A $\times$ [Fe] + B}.
\scriptsize
\begin{tabular}{ccccccccc} \hline \hline
Fit Parameter & $\displaystyle \frac{C_{33}}{C_{33}^{Fe-Poor}}$ & $\displaystyle \frac{C_{44}}{C_{44}^{Fe-Poor}}$ & $\displaystyle \frac{C_{55}}{C_{55}^{Fe-Poor}}$ & $\displaystyle \frac{C_{13}}{C_{13}^{Fe-Poor}}$ & $\displaystyle \frac{C_{15}}{C_{15}^{Fe-Poor}}$ & $\displaystyle \frac{C_{25}}{C_{25}^{Fe-Poor}}$ & $\displaystyle \frac{C_{35}}{C_{35}^{Fe-Poor}}$ & $\displaystyle \frac{C_{46}}{C_{46}^{Fe-Poor}}$ \\ \hline
A & -0.0167 & -0.0572 & -0.0345 & -0.0444 & -0.0249 & 0.0208 & -0.1060 & 0.0168 \\
B & 1.032 & 1.091 & 1.065 & 1.0747 & 1.062 & 0.991 & 1.150 & 0.995 \\ \hline \hline
\end{tabular}
\label{tab:Cij_vs_[Fe]_fits}
\end{table}
\subsubsection{Elastic Stability}
\begin{figure}
    \centering
    \includegraphics[scale = 0.45]{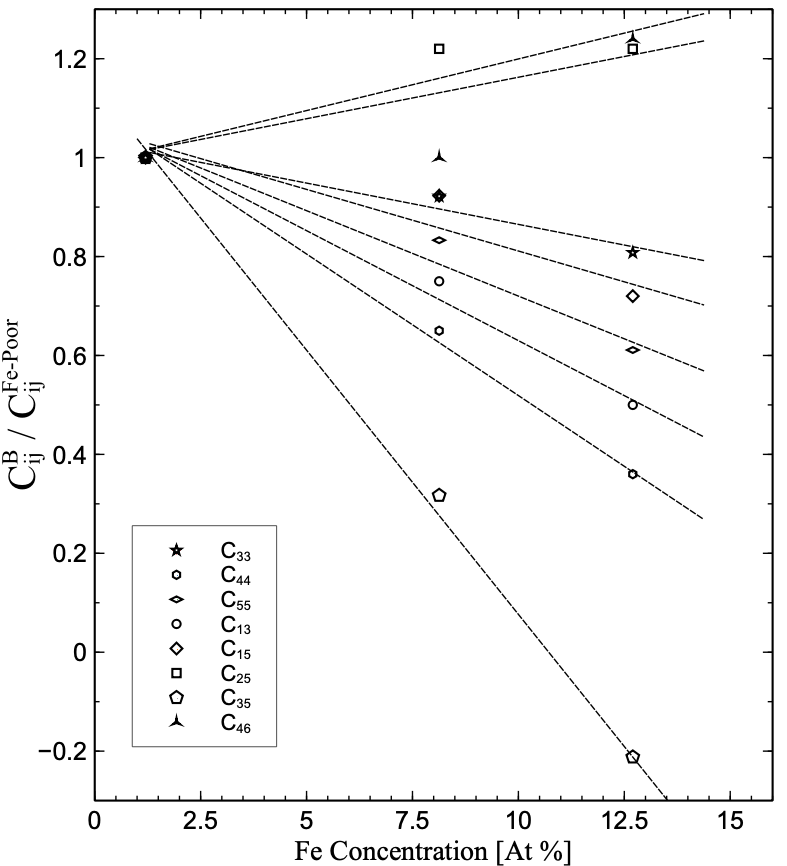}
    \caption{Selected elastic constants for biotite samples of the present work, each normalized to the corresponding constant for Fe-poor biotite (1.1\% Fe, 12.3\% Mg) ({\it i.e.}, $C^B_{ij}/C^{Fe-Poor}_{ij}$, where $ij =$ 33, 44, 55, 13, 15, 25, 35, 46), versus Fe concentration. Dashed lines - lines of best-fit to experimental data.}
    \label{fig:Cij_vs_Fe_Conc}
\end{figure}

\begin{figure}
\centering
\includegraphics[scale=0.4]{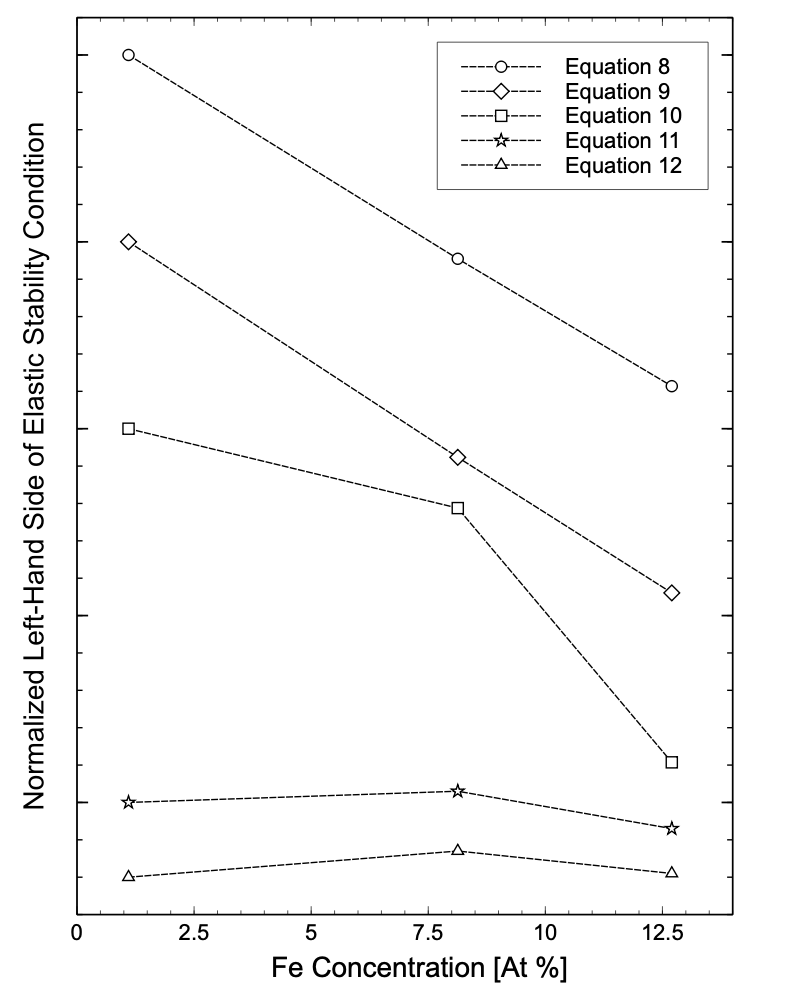}
  \caption{Left-hand side of elastic stability conditions (Eqs. 8-12) versus Fe concentration.  For each condition, the value of the left-hand side is normalized to that at lowest Fe concentration.  For example, for the condition given by Eq. \ref{stabcon1}, the left-most data point (at [Fe] = 1.1\%) has an ordinate of unity. The next point, at [Fe] = 8.1\%, is down 5 vertical “ticks” from the left-most point to an ordinate value of 0.50 and the third point, at the highest Fe concentration of 12.7\%, is down 9 vertical “ticks” from the left-most point to an ordinate value of 0.10.}
  \label{fig:stability}
\end{figure}

Fig.\,\ref{fig:stability} shows the value of the left-hand side of stability conditions Eqs.\,\ref{stabcon1}-\ref{stabcon5}, each normalized to the corresponding value at lowest Fe concentration, as a function of Fe concentration.  While the values of the left-hand sides of Eqs.\,\ref{stabcon4} and \ref{stabcon5} are nearly independent of Fe concentration, those of Eqs. \ref{stabcon1}-\ref{stabcon3} exhibit a dramatic decrease with increasing Fe concentration, being reduced, at the highest Fe concentration, to $\sim10$\% of the respective low-Fe concentration values.   Given that violation of one or more of the stability conditions results in an unstable crystal, the latter behaviour suggests that biotite becomes progressively less stable with increasing Fe content in the octahedral sheet. The same conclusion is reached when one considers the stability condition that the diagonal components of the elastic constants tensor be positive definite.  $C_{11}$, $C_{22}$, and $C_{66}$ are nearly independent of [Fe], but $C_{33}$, $C_{44}$, and $C_{55}$ show relatively large or very large decreases with increasing Fe concentration (see Table \ref{tab:elasticconstants} and Fig.\,\ref{fig:Cij_vs_Fe_Conc}).

The decrease in elastic stability with increasing Fe concentration is likely related to the accompanying increase in average octahedral cation radius.  In fact, synthetic trioctahedral micas of the form KR$_{3}^{2+}$AlSi$_{3}$O$_{10}$(OH)$_{2}$ with octahedral cation radius greater than that of Fe$^{2+}$ are not stable, the instability being attributed to the inability of the tetrahedral layer to further expand by rotation of tetrahedra, resulting in smaller tetrahedral layers on a larger octahedral layer \cite{haze1972}.  A second study shows that the stability is limited not only by tetrahedral rotation angle, but also by the absolute size of the octahedra \cite{tora1981}.  In the KMg$_{3}^{2+}$AlSi$_{3}$O$_{10}$(OH)$_{2}$ - KFe$_{3}^{2+}$AlSi$_{3}$O$_{10}$(OH)$_{2}$ system, the misfit between the tetrahedral and octahedral sheets decreases by substitution of large Fe$^{2+}$ or Mn$^{2+}$ ions for Mg$^{2+}$ ions. This induces variations of shared edge lengths in edge-shared octahedra which, in turn, increases the mutual repulsion between octahedral cations, resulting in a corresponding decrease in elastic stability \cite{tora1981}.  Previous studies have also shown that annite, an Fe-rich biotite, is much less stable than the Fe-poor biotite phlogopite \cite{eugs1962}.  Also consistent with this argument is that Mg silicates have, in general, a higher melting point than their Fe analogues \cite{gowe1957}. The results obtained in the present study are consistent with those of the above works and also show that the elastic stability exhibits a progressive decrease from Fe-poor (Mg-rich) biotite to Fe-rich (Mg-poor) biotite.

\subsubsection{Comparison to Theory}
Due to the similarity in compositions, the elastic constants of the Fe-poor (Mg-rich) biotite sample can be compared to those of phlogopite obtained from first-principles calculations based on density functional theory  \cite{chhe2014}.  As seen in Table \ref{tab:elasticconstants} and in Figure \ref{fig:Cefftrends}, most of the elastic constants for Fe-poor biotite are within $\leq10$\% of the corresponding values obtained for phlogopite using the generalized gradient approximation (GGA) or the local density approximation (LDA).  $C_{11}$, $C_{22}$, $C_{33}$, $C_{44}$, and $C_{55}$ show better agreement with values obtained using the GGA, while constants $C_{66}$, $C_{13}$, $C_{15}$, $C_{23}$, and $C_{46}$ compare very well with values obtained using the LDA.  Experimental values of $C_{25}$ and $C_{35}$ agree neither with the GGA nor the LDA values.  It is also interesting to note that the ratios $C_{22}/C_{11}$, $C_{55}/C_{44}$, and $C_{23}/C_{13}$ for phlogopite obtained from the GGA and LDA calculations are nearly identical to one another and, furthermore, are consistent with those for the Fe-poor biotite (see Table \ref{tab:elasticconstants}). Moreover, $C_{33}$, $C_{44}$, and $C_{55}$ for the Fe-poor biotite are smaller than the corresponding values for phlogopite obtained  using the GGA, while $C_{15}$ for the Fe-poor biotite is larger than the LDA-calculated value for phlogopite.  These behaviours ({\it i.e.}, $C^{Fe-poor}_{ii} \lesssim C^{Phlog}_{ii}$, $i = 3, 4, 5$, and $C^{Fe-poor}_{15} \gtrsim C^{Phlog}_{15}$) are as would be expected given the systematic dependence (strictly increasing or strictly decreasing) of these constants on Fe concentration as demonstrated in the present work, and the fact that elastic constants determined using the GGA and LDA tend to represent lower and upper bounds on the $C_{ij}$, respectively. We note that our results in Table \ref{tab:elasticconstants} also indicate that $C_{13}$, like $C_{33}$, $C_{44}$, and $C_{55}$, is a strictly decreasing function of [Fe] and while its behaviour does not mimic that of these other constants in the sense that $C^{Fe-poor}_{13} \nlesssim C^{Phlog}_{13}$ calculated using the GGA, the value of $C^{Phlog}_{13}$ obtained using the LDA is $\sim0.9\times C^{Fe-poor}_{13}$.

The first-principles calculations cited above yield an elastic constants tensor for trioctahedral phlogopite that is similar to that of dioctahedral muscovite, leading the authors to conclude that the elastic properties of micas are rather insensitive to octahedral site chemistry \cite{chhe2014}.  While it is true that the tensors are similar, the systematic changes with Fe concentration of several of the biotite elastic constants in the present work suggest that the octahedral site chemistry does in fact play an important role in determining the elastic properties of these micas. 

\subsubsection{Comparison to Muscovite}
The biotite elastic constants can be compared to those of muscovite determined in the present and previous studies (see Table \ref{tab:elasticconstants} and Fig.~\ref{fig:Cefftrends}) \cite{vaug1986,mcne1993}.  $C_{11}$, $C_{22}$, and $C_{66}$ values for biotites are very similar to those for muscovites.  This is not unexpected because, for both of these mica subgroups, the intralayer bonding is strong and the layers are composed of many of the same types of atoms arranged in a like fashion. $C_{23}$ also displays this behaviour, although the reason for its similarity to the corresponding value for muscovite is not obvious.  The values of $C_{33}$, $C_{44}$, and $C_{55}$ approach those of muscovite as Fe concentration decreases (or, equivalently, Mg concentration increases). The same general behaviour is observed for $C_{13}$.  It is difficult to make meaningful comparisons for $C_{15}$, $C_{25}$, $C_{35}$, and $C_{46}$ because of the very large variation in the values of these constants for muscovite.

The elastic anisotropy within the basal cleavage plane of biotite can be compared to that for muscovite via the ratios $C_{22}/C_{11}$ and $C_{55}/C_{44}$, both of which would be equal to unity if the basal plane displayed elastic isotropy. As Table \ref{tab:elasticconstants} shows, for all biotite and muscovite samples, $C_{22}/C_{11} \approx 1.0$.  In contrast, $C_{55}/C_{44}$ differs substantially from this value, being largest for Fe-rich biotite (2.5) and approaching values similar to those for muscovite as Fe content decreases. This result indicates that the basal plane anisotropy of Fe-rich biotite is strongest for the shear properties and is considerably higher than that of muscovite and decreases with decreasing Fe content in the octahedral sheet. \cite{vaug1986,mcne1993}

\section{\label{sec:Conclusion}Implications}
Brillouin light scattering spectroscopy was used to probe the elastic properties of natural biotite crystals with compositions quantified by electron probe microanalysis.  The systematic dependence of elastic wave velocities, elastic stiffness constants, elastic stability, and basal plane anisotropy on iron content suggests that biotite elasticity is a function of octahedral site chemistry and provides a means to estimate the elastic properties of most biotite compositions with known iron or magnesium content.  Moreover, the overall agreement between the experimentally-determined elastic constants of a Mg-rich biotite and those for phlogopite obtained from first-principles calculation based on density functional theory suggests that the latter approach holds promise in describing the elastic properties of biotites.

\section*{Acknowledgement}
\noindent
GTA acknowledges the support of the Natural Sciences and Engineering Research Council of Canada (NSERC) (RGPIN-2015-04306).

\vspace*{0.3cm}
\noindent
Additional data and code available upon request.






\bibliographystyle{elsarticle-num-names} 
\bibliography{bibliography}

\clearpage

\end{document}